\documentclass[english,preprint,nofootinbib]{revtex4}%
\usepackage[T1]{fontenc}
\usepackage[latin9]{inputenc}
\usepackage{amsmath}
\usepackage{amssymb}
\usepackage{amsfonts}
\usepackage{babel}
\usepackage{graphicx}%
\providecommand{\U}[1]{\protect\rule{.1in}{.1in}}
\makeatletter
\@ifundefined{textcolor}{}
{
\definecolor{BLACK}{gray}{0}
 \definecolor{WHITE}{gray}{1}
 \definecolor{RED}{rgb}{1,0,0}
 \definecolor{GREEN}{rgb}{0,1,0}
 \definecolor{BLUE}{rgb}{0,0,1}
 \definecolor{CYAN}{cmyk}{1,0,0,0}
 \definecolor{MAGENTA}{cmyk}{0,1,0,0}
 \definecolor{YELLOW}{cmyk}{0,0,1,0}
 }
\makeatother
\begin{document}
\title{Chemical potentials in three-dimensional higher spin anti-de Sitter gravity}
\author{Marc Henneaux$^{1,2}$, Alfredo Pérez$^{2}$, David Tempo$^{2}$, Ricardo
Troncoso$^{2,3}$}
\email{henneaux@ulb.ac.be,aperez@cecs.cl, tempo@cecs.cl, troncoso@cecs.cl}
\affiliation{$^{1}$Physique théorique et mathématique and International Solvay Institutes,
Université Libre de Bruxelles, Campus Plaine C.P.231, B-1050 Bruxelles, Belgium.}
\affiliation{$^{2}$Centro de Estudios Científicos (CECs), Casilla 1469, Valdivia, Chile}
\affiliation{$^{3}$Universidad Andrés Bello, Av. República 440, Santiago, Chile.}
\preprint{CECS-PHY-13/03}

\begin{abstract}
We indicate how to introduce chemical potentials for higher spin charges in
higher spin anti-de Sitter gravity in a manner that manifestly preserves the
original asymptotic $W$-symmetry. This is done by switching on a non-vanishing
component of the connection along the temporal (thermal) circles. We first
recall the procedure in the pure gravity case (no higher spin) where
the only \textquotedblleft chemical potentials\textquotedblright\ are the
temperature and the chemical potential associated with the angular momentum.
We then generalize to the higher spin case. We find that there is no tension
with the $W_{N}$ or $W_{\infty}$ asymptotic algebra, which is obviously
unchanged by the introduction of the chemical potentials. Our argument is not perturbative in the chemical potentials.

\end{abstract}
\maketitle

\section{Introduction}

Higher spin gauge theories in 3 spacetime dimensions
\cite{Vasiliev:1986qx,Blencowe, BBS,Vasiliev:1995dn}, which provide a useful
laboratory for understanding higher spin gauge theories in 4 and higher
dimensions \cite{VV0, VV1, VV2}, have attracted recently a considerable amount
of interest. One reason for this surge of activity is the rich asymptotic
structure displayed by the theory at infinity, where the $W$-algebras, or
their supersymmetric extensions in the graded case, emerge as asymptotic
symmetry algebras \cite{Henneaux-HS,Theisen-HS,Henneaux:2012ny}. This opens
the door to an investigation of holography with the powerful tools of
two-dimensional conformal field theory and representation theory of
$W$-algebras \cite{Gaberdiel:2010ar,Gaberdiel:2010pz,GH,ReviewHS5}.

As it is by now well known, anti-de Sitter gravity in 3 dimensions is
described by a $sl(2,%
\mathbb{R}
)\oplus sl(2,%
\mathbb{R}
)$ gauge theory \cite{Achucarro:1987vz,Witten:1988hc}. It was recognized in
\cite{Coussaert:1995zp} that the conditions expressing that the gravitational
field approaches at infinity the anti-de Sitter solution give, in the
Chern-Simons formulation, the conditions implementing the familiar Hamiltonian
reduction of the $sl(2,%
\mathbb{R}
)$-current algebra to the Virasoro algebra \cite{HR1,HR2,HR3,HR4}. This yields the gauge theory
derivation of the asymptotic Virasoro algebra and central charge first
obtained in \cite{Brown:1986nw} in the metric formulation.

The remarkable fact that the geometrical anti-de Sitter boundary conditions
implement the algebraic Hamiltonian reduction remains valid for simple and
extended supergravities \cite{Banados:1998pi,Henneaux:1999ib} and for higher
spin gauge theories \cite{Henneaux-HS,Theisen-HS,Henneaux:2012ny}.

Recently, exact black hole solutions supporting a non trivial higher spin
field have been obtained \cite{GK, GKMP,Ammon:2012wc} (see also \cite{CM}).
However, in spite of the simplicity of these black hole solutions, a suitable
characterization of their global charges and their entropy is a subject which
is not free of controversy, because there are tensions between various
approaches, which give different results (see \cite{deBoer:2013gz} for a lucid
discussion). We show in this paper that this tension is somewhat artificial
because it results from a non standard incorporation of the chemical
potentials that obscures the asymptotics and hence the correct definition of
the charges. Once the chemical potentials are properly introduced along the
lines indicated below, there is no difficulty with the asymptotics.
Our approach uses the Hamiltonian formalism, which provides a particularly
transparent analysis. It is not perturbative. 

Our paper is organized as follows. In the next section, we recall how the
chemical potentials are introduced in the metric formulation of pure gravity
in three dimensions and then translate the results in Chern-Simons terms. We
find that the chemical potentials appear through the temporal components of
the connection (along the thermal circles). This is in perfect agreement with
experience from four-dimensional gravity where the chemical potential for the
electric charge is well known to be associated with the zeroth component of the
electromagnetic vector potential in the
Reissner-Nordstr\"om solution. In Section \ref{Higher} we extend the
analysis to include higher spin charges and their chemical potentials. The
approach makes it obvious that non-vanishing chemical potentials do not change
the asymptotical properties because these potentials enter only the Lagrange
multipliers. Finally, we give comments and conclusions in Section
\ref{Conclusions}. We display a black hole solution that fulfills our
conditions. In a subsequent paper \cite{ToAppear,ToAppear2}, we shall further discuss
the asymptotics and the thermodynamics of this solution. It should be stressed
that our method agrees with the discussion of \cite{PTT1,PTT2}.

\section{Chemical potentials in the Chern-Simons formulation of pure anti-de
Sitter gravity}

\label{pure}

To begin with, we start with a discussion on how the temperature and the
chemical potential for the angular momentum enter in the $sl(2,%
\mathbb{R}
)\oplus sl(2,%
\mathbb{R}
)$ formulation of gravity. This simpler case illuminates the central points.  Similar considerations may be found in \cite{Banados:2012ue}.

\subsection{Metric formulation}

In the usual formulation of black hole thermodynamics, the temperature and the
chemical potential for the angular momentum do not enter the metric of the
black hole explicitly. They appear indirectly through the identifications
involving the imaginary time and the angle, which must be made to avoid a
singularity at the horizon in the Euclidean section. This means that the range
of the coordinates is not fixed but varies from one solution to another.

It is useful to have a description in which the range of the coordinates is
fixed once and for all. This can be achieved by redefining the time
coordinates $t\rightarrow\lambda t^{\prime}$ and $\theta\rightarrow
\theta^{\prime}=\theta+\omega t$, where $\lambda$ and $\omega$ are chosen such
that $t^{\prime}$ and $\theta^{\prime}$ have a constant range. This induces a
non trivial lapse and shift in the three-dimensional black hole solution
\cite{BTZ,BHTZ}, which reads (dropping primes on coordinates),
\begin{equation}
ds^{2}=-(N_{\infty})^{2}f^{2}dt^{2}+f^{-2}dr^{2}+r^{2}\left[  \left(
-\frac{J}{2r^{2}}N_{\infty}+N_{\infty}^{\theta}\right)  dt+d\theta\right]
^{2}\label{Metric}%
\end{equation}
with
\begin{equation}
f^{2}=\left(  \frac{r}{l}\right)  ^{2}-M+\frac{J^{2}}{4r^{2}}.\label{f}%
\end{equation}
If  one chooses the coordinates $t$ and $\theta$ such that $N_{\infty}%
=1$, $N_{\infty}^{\theta}=0$, then the ranges of the identifications in $t$
and $\theta$ depend on the solution. If one wants fixed ranges, one must therefore allow
for  $N_{\infty}$ and $N_{\infty}^{\theta}$ to vary. We impose that on the
Euclidean section $t\sim t+2\pi l$ and $\theta\sim\theta+2\pi$ (always).
The variables $N_{\infty}$ and $N_{\infty}^{\theta}$ are clearly related to the temperature
and the chemical potential for the angular momentum and will for this reason
be called \textquotedblleft the chemical potentials\textquotedblright. [We use quotation marks here because the temperature stands on a special footing but nevertherless it is convenient in what follows to include it among the standard chemical potentials.]

We shall from now on deal with 
the grand canonical ensemble, where the chemical potentials are held fixed to arbitrary values.  The appropriate variational principle
has then $N_{\infty}$ and $N_{\infty}^{\theta}$ fixed. One finds the value of
the conjugate variables, namely the mass $M$ and the angular momentum $J$
on-shell, by requiring the absence of singularity in the Euclidean section at
the horizon, which imposes in particular that $N^{\theta}= - \frac{J}{2 r^{2}}
N_{\infty}+ N^{\theta}_{\infty}$ should vanish at the horizon.

When $N_{\infty}\not =1$ and $N_{\infty}^{\theta}\not =0$, the metric does not
fulfill at infinity the boundary conditions of \cite{Brown:1986nw}, which,
from the present perspective, would correspond to fixed $\beta=\frac{1}{2\pi
l}$ and zero chemical potential for the angular momentum. However, it is very
easy to translate these boundary conditions to generic values of the chemical
potentials, just like it is very easy to translate the asymptotic flat
boundary conditions written in cartesian coordinates to spherical coordinates
through the appropriate coordinate transformation. The asymptotic symmetry is
of course the same. When the chemical potentials are introduced, one should
not talk about a relaxation of the boundary conditions, but rather of a
(straightforward in this case) extension of the formalism to cover different
values of the (held fixed) chemical potentials.

The only case where the metric is not asymptotically AdS is when $N_{\infty}=
0$, which corresponds to the infinite temperature limit and to a degenerate
metric ($\det g = 0$). We shall not consider this case in this paper.

\subsection{Connection formulation}

How do the chemical potentials enter the Chern-Simons connection? We claim
that they appear as additional contributions to the thermal circles around the
horizon ($dt$ contributions to the connection), explicitly (after the
$r$-dependent gauge transformation of \cite{Coussaert:1995zp} has been
performed to eliminate the $r$-dependence to leading order):
\begin{equation}
a^{\pm}=\pm\left(  L_{\pm1}^{\pm}-\frac{2\pi}{k}\mathcal{L}_{\pm}L_{\mp1}%
^{\pm}\right)  dx^{\pm}\pm\frac{1}{l}\Lambda^{\pm}(\nu_{\pm}%
)dt\ ,\label{ChemPot}%
\end{equation}%
\[
\Lambda^{\pm}\left(  \nu_{\pm}\right)  =\nu_{\pm}L_{\pm1}^{\pm}-\frac{2\pi}%
{k}\nu_{\pm}\mathcal{L}_{\pm}L_{\mp1}^{\pm}\ ,
\]
(asymptotically) where $\nu^{\pm}$ are constants and called the chemical
potentials of the Chern-Simons formulation\footnote{In our conventions the
level is given by $k=\frac{l}{4G}=2 l$.}. Indeed, with constant
$\mathcal{L}_{\pm}$'s, the metric corresponding to (\ref{ChemPot}) is
(\ref{Metric}) with
\begin{align}
&  \left(  N_{\infty}\right)  ^{2}=\frac{1}{4}\left(  \nu^{+}+\nu
^{-}+2\right)  ^{2}\\
&  N_{\infty}^{\theta}=\frac{\nu^{+}-\nu^{-}}{2l}%
\end{align}
and
\begin{equation}
M=\frac{2\pi}{l}\left(  \mathcal{L}_{+}+\mathcal{L}_{-}\right)  \;\;\;\;J=2\pi
\left(  \mathcal{L}_{+}-\mathcal{L}_{-}\right)  .
\end{equation}

\subsection{Asymptotic Analysis}

We now show that the introduction of the chemical potentials does not modify the
Virasoro asymptotics. This is in fact direct, and physically mandatory, but we
provide an explicit argument since some confusion arose in the spin-3 case.

The discussion is most transparent in the Hamiltonian formalism. On a slice
$t=\text{const}$, say the initial slice $t=0$, the connection is
asymptotically given by
\begin{equation}
a^{\pm}(t=0)=\left(  L_{\pm1}^{\pm}-\frac{2\pi}{k}\mathcal{L}_{\pm}L_{\mp
1}^{\pm}\right)  d\theta\label{Asymp}%
\end{equation}
The gauge transformations that preserve this form of the connection are
asymptotically parametrized by a gauge parameter that takes the form
\begin{equation}
\Lambda^{\pm}\left(  \varepsilon_{\pm}\right)  =\varepsilon_{\pm}L_{\pm1}%
^{\pm}\mp\varepsilon_{\pm}^{\prime}L_{0}^{\pm}+\frac{1}{2}\left(
\varepsilon_{\pm}^{\prime\prime}-\frac{4\pi}{k}\varepsilon_{\pm}%
\mathcal{L}_{\pm}\right)  L_{\mp1}^{\pm} \label{Lambda0}%
\end{equation}
where $\varepsilon_{\pm}$ are at this stage arbitrary functions of $\theta$
\emph{and also} of the slice under consideration, i.e., $t$, since one can make
independent gauge transformations that preserve (\ref{Asymp}) on each slice.
Here, prime denotes the derivative with respect to $\theta$.

The motion from one slice to the next is a gauge transformation parametrized
by the Lagrange multiplier $a_{0}^{\pm}$ associated with the Chern-Simons
Gauss constraint. To preserve the asymptotic form (\ref{Asymp}), $a_{0}^{\pm}$
should be of the form (\ref{Lambda0}). The choice of the Lagrange multiplier
which is made when the chemical potentials are not switched on is simply
$\varepsilon_{\pm}=1$, so that $a_{0}^{\pm}=\pm a_{\theta}^{\pm}$. The
equations of motion imply that the fields are chiral with $\pm$ chiralities,
and asymptotically given by (\ref{ChemPot}) with $\nu_{\pm}=0$.

The choice of Lagrange multipliers which is made when the chemical potentials
are switched on is $\varepsilon_{\pm}=1+\nu_{\pm}$ yielding now (\ref{ChemPot}%
) with $\nu_{\pm}$ non zero when one integrates the equations. It is
immediate, by very construction, that:

\begin{itemize}
\item The asymptotic symmetry algebra is the conformal algebra since the
connection obeys (\ref{Asymp}) on all slices (the Lagrange multipliers are
taken in the allowed class of gauge parameters).

\item The $\mathcal{L}_{\pm}$ fulfill Virasoro algebra with the same central
charge independently of $\nu_{\pm}$ since they depend only on the canonical
variables and not on the Lagrange multipliers.
\end{itemize}

To close this subsection, we note that the introduction of the chemical
potentials through the temporal components of the connection (i.e., the
components along the thermal circles) is in fact familiar from the
thermodynamics of Reissner-Nordstr\"om black holes, where the chemical potential for the electric charge is introduced through the temporal component $A_0$ of the electromagnetic vector potential. $A_0$  is the Lagrange multiplier for Gauss' law. This procedure actually guarantees
the interpretation of the Lagrange multipliers as chemical potentials. Indeed, quite generally, the
Lagrange multipliers $\lambda^A$ enter the action as $- \int d^d x \lambda^A {\mathcal G}_A -  \lambda^A_\infty q_A $ where ${\mathcal G}_A$ are the constraints (which
vanish on-shell) and $q_A$ the charges, which are given by the surface terms at
infinity that must accompany the bulk constraints to make the variational
principle well defined \cite{Regge:1974zd}.  On shell, this sum reduces to $-\lambda^A_\infty q_A $.

\subsection{Some comments}

We thus see that it is rather straightforward to handle the chemical
potentials in the Chern-Simons formulation. It has been proposed in the
literature to include them not along the thermal circle, but along the
conjugate null directions, as
\begin{equation}
a^{\pm}=\pm\left(  L_{\pm1}^{\pm}-\frac{2\pi}{k}\mathcal{L}_{\pm}L_{\mp1}%
^{\pm}\right)  dx^{\pm}\pm\Lambda^{\pm}(\nu_{\pm})dx^{\mp}%
\ .\label{ChemPotBis}%
\end{equation}

Now, this is more than just a choice of the Lagrange multiplier. One sees that
this choice has also the effect of modifying the phase space variables
$a_{\theta}^{\pm}$ as
\[
a^{\pm}(t= \textrm{const})=\left(  (1-\nu_{\pm})L_{\pm1}^{\pm}-\frac{2\pi(1-\nu_{\pm})}%
{k}\mathcal{L}_{\pm}L_{\mp1}^{\pm}\right)  d\theta,   
\]
which is not any more of the requested asymptotic form (\ref{Asymp}). However, even though
not of the requested asymptotic form, one can bring the spatial connection to
it by redefinitions, so that the \textquotedblleft penalty\textquotedblright%
\ paid is not very high: the formulas need only direct, although somewhat
awkward, adjustments. As we shall see, this is not the case for the higher
spin charges, where this different approach destroys the asymptotics.

\section{Higher spin Chemical Potentials}

\label{Higher}

We now turn to the higher spin case. For definite, we consider the theory
based on $sl\left(  3,%
\mathbb{R}
\right)  \oplus sl\left(  3,%
\mathbb{R}
\right)  $, which contains a spin 3 field in addition to the spin 2 one, and
which illustrates the main points.

\subsection{Direct approach}

The asymptotic form of the connections when the chemical potentials are not
included in them read \cite{Henneaux-HS,Theisen-HS}, after the $r$-dependent
gauge transformation of \cite{Coussaert:1995zp} has been performed,
\begin{equation}
a^{\pm}=\pm\left(  L_{\pm1}^{\pm}-\frac{2\pi}{k}\mathcal{L}_{\pm}L_{\mp1}%
^{\pm}-\frac{\pi}{2k}\mathcal{W}_{\pm}W_{\mp2}^{\pm}\right)  dx^{\pm}\ .
\label{StandardW3}%
\end{equation}
So, on a slice $t = \text{const}$, the spatial connection (which contains the
reduced phase space variables) takes the asymptotic form
\begin{equation}
a^{\pm}(t=\textrm{const})=\left(  L_{\pm1}^{\pm}-\frac{2\pi}{k}\mathcal{L}_{\pm}L_{\mp
1}^{\pm}-\frac{\pi}{2k}\mathcal{W}_{\pm}W_{\mp2}^{\pm}\right)  d\theta\ .
\label{AsymptoticW3}%
\end{equation}

The gauge transformations that leave this asymptotic form invariant are
\cite{Henneaux-HS,Theisen-HS}
\begin{align}
\Lambda^{\pm}\left(  \varepsilon_{\pm},\chi_{\pm}\right)   &  =\varepsilon
_{\pm}L_{\pm1}^{\pm}+\chi_{\pm}W_{\pm2}^{\pm}\mp\varepsilon_{\pm}^{\prime
}L_{0}^{\pm}\mp\chi_{\pm}^{\prime}W_{\pm1}^{\pm}+\frac{1}{2}\left(
\varepsilon_{\pm}^{\prime\prime}-\frac{4\pi}{k}\varepsilon_{\pm}%
\mathcal{L}_{\pm}+\frac{8\pi}{k}\mathcal{W}_{\pm}\chi_{\pm}\right)  L_{\mp
1}^{\pm}\nonumber\\
&  -\left(  \frac{\pi}{2k}\mathcal{W}_{\pm}\varepsilon_{\pm}+\frac{7\pi}%
{6k}\mathcal{L}_{\pm}^{\prime}\chi_{\pm}^{\prime}+\frac{\pi}{3k}\chi_{\pm
}\mathcal{L}_{\pm}^{\prime\prime}+\frac{4\pi}{3k}\mathcal{L}_{\pm}\chi_{\pm
}^{\prime\prime}\right.  \left.  -\frac{4\pi^{2}}{k^{2}}\mathcal{L}_{\pm}%
^{2}\chi_{\pm}-\frac{1}{24}\chi_{\pm}^{\prime\prime\prime\prime}\right)
W_{\mp2}^{\pm}\nonumber\\
&  +\frac{1}{2}\left(  \chi_{\pm}^{\prime\prime}-\frac{8\pi}{k}\mathcal{L}%
_{\pm}\chi_{\pm}\right)  W_{0}^{\pm}\mp\frac{1}{6}\left(  \chi_{\pm}%
^{\prime\prime\prime}-\frac{8\pi}{k}\chi_{\pm}\mathcal{L}_{\pm}^{\prime}%
-\frac{20\pi}{k}\mathcal{L}_{\pm}\chi_{\pm}^{\prime}\right)  W_{\mp1}^{\pm}\ ,
\label{LambdaW3}%
\end{align}
where $\varepsilon_{\pm}$, $\chi_{\pm}$ are arbitrary functions of $\theta$
(on each slide). Furthermore, the functions $\mathcal{L}_{\pm}$ and
$\mathcal{W}_{\pm}$ of the canonical variables appearing in the asymptotic
form of the connection transform as \cite{Henneaux-HS,Theisen-HS}
\begin{align}
\delta\mathcal{L}_{\pm}  &  =\varepsilon_{\pm}\mathcal{L}_{\pm}^{\prime
}+2\mathcal{L}_{\pm}\varepsilon_{\pm}^{\prime}-\frac{k}{4\pi}\varepsilon_{\pm
}^{\prime\prime\prime}-2\chi_{\pm}\mathcal{W}_{\pm}^{\prime}-3\mathcal{W}%
_{\pm}\chi_{\pm}^{\prime}\ ,\label{DeltaLW3}\\
\delta\mathcal{W}_{\pm}  &  =\varepsilon_{\pm}\mathcal{W}_{\pm}^{\prime
}+3\mathcal{W}_{\pm}\varepsilon_{\pm}^{\prime}-\frac{64\pi}{3k}\mathcal{L}%
_{\pm}^{2}\chi_{\pm}^{\prime}+3\chi_{\pm}^{\prime}\mathcal{L}_{\pm}%
^{\prime\prime}+5\mathcal{L}_{\pm}^{\prime}\chi_{\pm}^{\prime\prime}+\frac
{2}{3}\chi_{\pm}\mathcal{L}_{\pm}^{\prime\prime\prime}-\frac{k}{12\pi}%
\chi_{\pm}^{\prime\prime\prime\prime\prime}\nonumber\\
&  -\frac{64\pi}{3k}\left(  \chi_{\pm}\mathcal{L}_{\pm}^{\prime}-\frac
{5k}{32\pi}\chi_{\pm}^{\prime\prime\prime}\right)  \mathcal{L}_{\pm}\ ,
\label{DeltaWW3}%
\end{align}
leading to the $W_{3}$ algebra (with central charge).

As we explained above, the Lagrange multipliers $a_{0}^{\pm}$ must preserve
the boundary conditions and so must be of the form (\ref{LambdaW3}). Now, in
the case where the chemical potentials are not incorporated in the connection
\cite{Henneaux-HS,Theisen-HS}, one simply takes $\varepsilon_{\pm}=1$,
$\chi_{\pm}=0$ and one finds that on-shell, the connections are chiral and
take the asymptotic form (\ref{AsymptoticW3}).

If one wants to include the chemical potentials in the connection, one should,
as explained above, allow for extra terms in $a_{0}^{\pm}$. The temporal
component of the connection reads now
\begin{equation}
a_{0}^{\pm}=\pm\left(  L_{\pm1}^{\pm}-\frac{2\pi}{k}\mathcal{L}_{\pm}L_{\mp
1}^{\pm}-\frac{\pi}{2k}\mathcal{W}_{\pm}W_{\mp2}^{\pm}\right)  \frac{dt}{l}%
\pm\frac{1}{l}\Lambda^{\pm}(\nu_{\pm},\mu_{\pm})dt\ ,\label{AsympWithChemical}%
\end{equation}
where $\nu_{\pm}$ and $\mu_{\pm}$ are \emph{constants} and are the respective
chemical potentials for the spin-2 part and the spin-3 part. The connection is
therefore
\begin{equation}
a^{\pm}=\pm\left(  L_{\pm1}^{\pm}-\frac{2\pi}{k}\mathcal{L}_{\pm}L_{\mp1}%
^{\pm}-\frac{\pi}{2k}\mathcal{W}_{\pm}W_{\mp2}^{\pm}\right)  dx^{\pm}\pm
\frac{1}{l}\Lambda^{\pm}(\nu_{\pm},\mu_{\pm})dt.\label{AsympWithChemicalmn}%
\end{equation}
Note that this is not really a relaxed set of boundary conditions since the
spatial part of the connection is unchanged. Only the Lagrange parameters are
modified. This is an extension of the formalism that incorporates the chemical potentials.

Just as in the pure gravity case, it is obvious that this choice is such that:

\begin{itemize}
\item The asymptotic symmetry algebra is the conformal $W_{3}$ algebra since
the connection obeys (\ref{AsymptoticW3}) on all slices (the Lagrange
multipliers are taken in the allowed class of gauge parameters).

\item The $\mathcal{L}_{\pm}$, $\mathcal{W}_{\pm}$ fulfill in the
Poisson-Dirac bracket the $W_{3}$ algebra of \cite{Henneaux-HS,Theisen-HS}
with the same central charge independently of the chemical potentials since
these generators depend only on the canonical variables and not on the
Lagrange multipliers.
\end{itemize}

\subsection{Comments}

If one were to introduce the chemical potentials through extra non-vanishing
components of the connection not along the thermal circles but along the
conjugate timelike directions, one would run into serious difficulties.
Indeed, if one were to impose asymptotically
\begin{equation}
a^{\pm}=\pm\left(  L_{\pm1}^{\pm}-\frac{2\pi}{k}\mathcal{L}_{\pm}L_{\mp1}%
^{\pm}-\frac{\pi}{2k}\mathcal{W}_{\pm}W_{\mp2}^{\pm}\right)  dx^{\pm}%
\pm\Lambda^{\pm}\left(  \nu_{\pm},\mu_{\pm}\right)  dx^{\mp},
\end{equation}
one would modify the spatial connection in a way incompatible with the $W_{3}$
symmetry since the terms proportional to the chemical potentials $\mu_{\pm}$
for the spin-3 charge enter $a_{\theta}^{\pm}$ multiplied by Lie algebra
generators that are not highest (lowest) weight states and hence are not
compatible with the asymptotic conditions (\ref{AsymptoticW3}) implementing the Hamiltonian reduction of the
$sl(3)$ current algebra to the $W_{3}$ algebra,
\begin{align}
a^{\pm}(t=\textrm{const}) &  =\left(  L_{\pm1}^{\pm}-\frac{2\pi}{k}\mathcal{L}_{\pm}%
L_{\mp1}^{\pm}-\frac{\pi}{2k}\mathcal{W}_{\pm}W_{\mp2}^{\pm}\right)
d\theta+(\nu_{\pm}L_{\pm1}^{\pm}+\mu_{\pm}W_{\pm2}^{\pm})d\theta\nonumber\\
&  +\left[  \frac{1}{2}\left(  -\frac{4\pi}{k}\nu_{\pm}\mathcal{L}_{\pm}%
+\frac{8\pi}{k}\mathcal{W}_{\pm}\mu_{\pm}\right)  L_{\mp1}^{\pm}-\left(
\frac{\pi}{2k}\mathcal{W}_{\pm}\nu_{\pm}-\frac{4\pi^{2}}{k^{2}}\mathcal{L}%
_{\pm}^{2}\mu_{\pm}\right)  W_{\mp2}^{\pm}\right]  d\theta\nonumber\\
&  -\frac{4\pi}{k}\mathcal{L}_{\pm}\mu_{\pm}W_{0}^{\pm}d\theta.
\end{align}
The \textquotedblleft offending terms\textquotedblright\ are absent when the
chemical potentials $\mu_{\pm}$ are zero (although rescalings are still needed
in that case), but present otherwise.

In fact, when the chemical potentials $\mu_{\pm}$ are non zero, the full
asymptotic asymptotic symmetry at infinity, i.e., the set of \emph{all} gauge
transformations preserving the asymptotic form of $a_{\theta}$ is the algebra
$W_{3}^{2}$ corresponding to the other non trivial embedding of $sl\left(  2,%
\mathbb{R}
\right)  $ into $sl\left(  3,%
\mathbb{R}
\right)  $ \cite{ToAppear,ToAppear2}. In the enveloping algebra of $W_{3}^{2}$, one may try to pick out
a $W_{3}$ algebra, for instance by requiring analyticity in $\mu$.
Perturbative efforts in that direction may be found in \cite{Compere:2013gja}
where the equations have been analyzed to  finite order  $O(\mu^4)$.

\section{Conclusions}

\label{Conclusions}

We have shown in this paper how to incorporate the chemical potentials
associated with higher spin charges in higher spin three-dimensional gravity.
Although we considered only the case of $sl\left(  3,%
\mathbb{R}
\right)  $, it is clear that our method extends straightforward to any
$sl\left(  N,%
\mathbb{R}
\right)  $ (with any non trivial embedding of $sl\left(  2,%
\mathbb{R}
\right)  $ and even to infinite-dimensional higher-spin algebras).  As here, the asymptotic symmetry algebra is obviously unchanged when the chemical potentials are switched on.

The method is straightforward because the chemical potentials enter only the
temporal components of the connection, which are Lagrange multipliers. The
canonical variables - and hence the canonical generators of the symmetry at
infinity - are unaffected. In that sense, the boundary conditions with
chemical potentials included are not true relaxations of the original boundary
conditions. True relaxations would be relaxations on the behavior of the
canonical variables and not just the Lagrange multipliers. The Hamiltonian formalism makes the
analysis particularly transparent and direct. The argument is non-perturbative and exact in the chemical potentials.

As we pointed out, the introduction of the chemical potentials through the
temporal (i.e., along the thermal circles) components of the connection is in
fact familiar from the thermodynamics of Reissner-Nordstr\"om black holes. The
thermodynamical significance of the Lagrange multipliers as chemical
potentials is  immediate given the structure of the action, where the
constraints and the accompanying charges (given by surface integrals at
infinity) are multiplied by the Lagrange multipliers. It is very satisfying
that what works in four dimensions also works in three.

In a future paper \cite{ToAppear,ToAppear2}, we shall provide more insight on the
analysis by  investigating the static black hole solution endowed with a
spin-$3$-field
\begin{align}
a^{\pm} &  =\left(  \pm L_{\pm1}^{\pm}\mp\frac{2\pi}{k}\mathcal{L}L_{\mp
1}^{\pm}-\frac{\pi}{2k}\mathcal{W}W_{\mp2}^{\pm}\right)  dx^{\pm
}\label{BH-connection-static}\\
&  +\left(  \mu W_{\pm2}^{\pm}-\frac{4\pi}{k}\mu\mathcal{L}W_{0}^{\pm}\pm\nu
L_{\pm1}^{\pm}\pm\frac{2\pi}{k}\left[  2\mu\mathcal{W}-\nu\mathcal{L}\right]
L_{\mp1}^{\pm}+\frac{\pi}{2k}\left[  \frac{8\pi}{k}\mu\mathcal{L}^{2}%
-\nu\mathcal{W}\right]  W_{\mp2}^{\pm}\right)  \frac{dt}{l}\ ,\nonumber
\end{align}
where $\mathcal{L}$, $\mathcal{W}$, $\mu$, $\nu$ are integration constants. We
shall confirm through the study of this specific ``$W_3$-black hole"  that there is no
tension between the holographic and canonical approaches. The analysis of the
thermodynamics and the conformal properties is direct, because the
coefficients ${\mathcal{L}}$ and ${\mathcal{W}}$ in the connection really mean
what they are, namely, the generators of the $W_{3}$ algebra, without needing
translation through a dictionary.

We shall also investigate the rotating solution, which is given by (\ref{AsympWithChemicalmn}) with
$\mathcal{L}_{\pm}$, $\mathcal{W}_{\pm}$, $\nu_{\pm}$ and $\mu_{\pm}$ fixed to
constants, which we write by displaying explicitly the spatial and temporal components%
\begin{align}
a^{\pm} &  =\left(  L_{\pm1}^{\pm}-\frac{2\pi}{k}\mathcal{L}_{\pm}L_{\mp
1}^{\pm}-\frac{\pi}{2k}\mathcal{W}_{\pm}W_{\mp2}^{\pm}\right)  d\theta\pm
\frac{1}{l}\left[  \left(  1+\nu_{\pm}\right)  L_{\pm1}^{\pm}+\mu_{\pm}%
W_{\pm2}^{\pm}-\frac{4\pi}{k}\mu_{\pm}\mathcal{L}_{\pm}W_{0}^{\pm}\right.
\nonumber\\
&  \left.  -\frac{2\pi}{k}\left(  \left(  1+\nu_{\pm}\right)  \mathcal{L}%
_{\pm}-2\mu_{\pm}\mathcal{W}_{\pm}\right)  L_{\mp1}^{\pm}-\frac{\pi}%
{2k}\left(  \left(  1+\nu_{\pm}\right)  \mathcal{W}_{\pm}-\frac{8\pi}{k}%
\mu_{\pm}\mathcal{L}_{\pm}^{2}\right)  W_{\mp2}^{\pm}\right]  dt\ ,
\end{align}
as well as the analog ``$W^{(2)}_3$-black holes".

Perhaps the following question can be asked as a final note. We have seen that
the chemical potentials $\nu_{\pm}$ of the gravitational sector can be
absorbed through a definition $t\rightarrow\alpha t$, $\theta\rightarrow
\theta+\omega t$ of the coordinates, at the price of having new ranges for the
new coordinates. It would be interesting to see if analogous absorptions of
the higher spin chemical potentials could take place by suitable redefinitions
in the yet-to-be-found geometry incorporating the higher spin fields.

\vspace{.3cm}
\noindent
Note added: In the interesting paper \cite{Compere:2013nba}, devoted to the higher spin black hole solutions of \cite{GK}, it is advocated that the chemical potentials should be defined in that case in terms of the components of the connection along the thermal circles.  This work relies on  the approach of \cite{Compere:2013gja}, where the  $W_3$-symmetry was  proved to be present (perturbatively,   to order $O(\mu^4)$).  We are grateful to the authors of \cite{Compere:2013nba} for drawing our attention on their work.

\acknowledgments We thank C. Martínez and C. Troessaert for helpful
discussions. M.H. thanks the Alexander von Humboldt Foundation for a Humboldt
Research Award. The work of M.H. is partially supported by the ERC through
the \textquotedblleft SyDuGraM\textquotedblright\ Advanced Grant, by IISN -
Belgium (conventions 4.4511.06 and 4.4514.08) and by the \textquotedblleft
Communauté Française de Belgique\textquotedblright\ through the ARC program.
The work of A.P., D.T. and R.T. is partially funded by the Fondecyt grants
N${^{\circ}}$ 1130658, 1121031, 3110122, 3110141. The Centro de Estudios
Científicos (CECS) is funded by the Chilean Government through the Centers of
Excellence Base Financing Program of Conicyt.

\end{document}